# An RF-source-free microwave photonic radar with an optically injected semiconductor laser for high-resolution detection and imaging

PEI ZHOU,[1,2,3,†] RENHENG ZHANG,[1,2,†] NIANQIANG LI,[1,2,4] ZHIDONG JIANG,[1,2] AND SHILONG PAN[3,5]

[1]*School of Optoelectronic Science and Engineering & Collaborative Innovation Center of Suzhou Nano Science and Technology, Soochow University, Suzhou 215006, China*
[2]*Key Lab of Advanced Optical Manufacturing Technologies of Jiangsu Province & Key Lab of Modern Optical Technologies of Education Ministry of China, Soochow University, Suzhou 215006, China*
[3]*Key Laboratory of Radar Imaging and Microwave Photonics, Ministry of Education, Nanjing University of Aeronautics and Astronautics, Nanjing 210016, China*
[4]*e-mail: nli@suda.edu.cn*
[5]*e-mail: pans@nuaa.edu.cn*

**This paper presents a novel microwave photonic (MWP) radar scheme that is capable of optically generating and processing broadband linear frequency-modulated (LFM) microwave signals without using any radio-frequency (RF) sources. In the transmitter, a broadband LFM microwave signal is generated by controlling the period-one (P1) oscillation of an optically injected semiconductor laser. After targets reflection, photonic de-chirping is implemented based on a dual-drive Mach-Zehnder modulator (DMZM), which is followed by a low-speed analog-to-digital converter (ADC) and digital signal processer (DSP) to reconstruct target information. Without the limitations of external RF sources, the proposed radar has an ultra-flexible tunability, and the main operating parameters are adjustable, including central frequency, bandwidth, frequency band, and temporal period. In the experiment, a fully photonics-based Ku-band radar with a bandwidth of 4 GHz is established for high-resolution detection and inverse synthetic aperture radar (ISAR) imaging. Results show that a high range resolution reaching ~1.88 cm, and a two-dimensional (2D) imaging resolution as high as ~1.88 cm × ~2.00 cm are achieved with a sampling rate of 100 MSa/s in the receiver. The flexible tunability of the radar is also experimentally investigated. The proposed radar scheme features low cost, simple structure, and high reconfigurability, which, hopefully, is to be used in future multifunction adaptive and miniaturized radars.**

## 1. INTRODUCTION

Compared with optical sensors, e.g., lidars, radars are more attractive in terms of the all-time and all-weather operation capability [1]. Fast, high-resolution detection and imaging of objects by radar is highly desired in many emerging applications, such as unmanned aerial vehicles (UAVs) and autonomous vehicles. In order to achieve high-speed and high-resolution target detection, a radio-frequency (RF) radar is required to have both a broadband transmit signal and a fast signal processing ability [2]. In conventional electrical radars, the transmit signal generated by purely electrical techniques suffers from limited frequency and bandwidth. Although the bandwidth can be increased by multiple stages of frequency conversion, the performance of the transmit signal in the transmitter as well as the analog-to-digital converters (ADCs) in the receiver would deteriorate rapidly with the increasement of the operating bandwidth. Therefore, the resolution and processing speed of the radar are restricted. To deal with these problems associated with electrical methods, numerous microwave photonic (MWP) radar schemes have been proposed [3, 4]. In 2014, P. Ghelfi *et al.* have demonstrated the first MWP radar, in which a mode-locked laser is employed to realize both signal generation in the transmitter and optical sampling in the receiver [5]. However, its operating bandwidth and ranging resolution are restricted by the small pulse-repetition frequency. To deal with this problem, different types of broadband linear frequency-modulated (LFM) signal generators have been utilized in MWP radars, including optical wavelength-to-frequency mapping [6], photonic digital-to-analog convertor [7], microwave photonic frequency multiplication [8], and so on. Among these schemes to realize broadband MWP radars, a major method is taking advantage of microwave photonic frequency multiplication for broadband signal generation in the transmitter, while microwave photonic frequency mixing for broadband photonic de-chirping in the receiver [8-12]. Thanks to the bandwidth compression property brought by broadband photonic de-chirping, fast and even real-time radar signal processing is possible. In 2017, a real-time and high-resolution MWP imaging radar was reported with an instantaneous bandwidth of 8 GHz and a two-dimensional (2D) imaging resolution of 2 cm × 2 cm [9]. After that, multiple-input-multiple-output (MIMO), multiband, and phased array MWP radars were also successfully demonstrated [13-18]. However, an intermediate-frequency linear frequency-modulated (IF-LFM) signal is required for those MWP radars based on photonic frequency multiplying. For example, in [12], an IF-LFM signal centered at 11 GHz with a bandwidth of 4 GHz is adopted for photonic frequency doubling, which inevitably increases the complexity and cost of the system. Besides, the electrical IF-LFM signal usually operates efficiently only in a particular frequency band with specific bandwidth, which would restrain the reconfigurable capabilities of radars. Recently, X. Zhang *et al*. have demonstrated an RF-source-free MWP radar based on a Fourier domain mode locking optoelectronic oscillator (FDML-OEO) [19]. The main drawback is the generated LFM signal has limited linearity in the transmitter, leading to poor performance of the photonic de-chirping process in the receiver. As a consequence, complicated compensation processing is necessary to obtain target information, which definitely slows down the processing accuracy and speed of the radar.

In recent years, a novel approach to generating microwave signals based on an optically injected semiconductor laser (OISL) has been proposed and received increasing attentions [20]. By taking advantage of the period-one (P1) dynamics of an OISL, photonic microwave generation with a tunable frequency from a few to over 100 GHz has been achieved without the requirements of any external RF sources [21]. In addition to tunable microwave signals [22-25], P1 oscillations of an OISL have also been applied for generating microwave frequency combs [26], optical frequency combs [27, 28], optical pulses [29], and triangular pulses [30]. In our prior work, we have proposed and demonstrated the photonic generation of broadband LFM signal by dynamically controlling the injection parameters of an OISL [31]. More importantly, the OISL-based LFM signal generator features ultra-flexible tunability, whose parameters, including central frequency, bandwidth, temporal duration, frequency band, and duty cycle, are easily adjustable [32].

In order to improve the spectral purity of the generated LFM signals, methods of optical feedback, optoelectronic feedback have been demonstrated [33-35]. Furthermore, photonic generation of other useful radar waveforms has also been reported based on the P1 dynamics of an OISL, such as frequency-hopping sequences [36], nonlinear frequency-modulated (NLFM) microwave waveforms [37], and dual-chirp LFM waveforms [38, 39].

In this paper, we put forward a new photonics-based radar with an optically injected semiconductor laser without the requirement of RF sources, and high-resolution ranging [40], as well as high-resolution imaging are realized. In the transmitter, a broadband LFM signal is generated by the controlled P1 oscillation of an OISL. In the receiver, photonic de-chirping and a low-speed ADC are used to realize fast signal processing. In the experimental demonstration, a fully photonics-based inverse synthetic aperture radar (ISAR) system with a bandwidth of 4 GHz is established, a high range-resolution reaching ~1.88 cm, and a 2D imaging resolution as high as ~1.88 cm × ~2.00 cm are achieved with a sampling rate of 100 MSa/s. Besides, the flexible tunability of the radar is also investigated by adjusting the central frequency, bandwidth, frequency band, and temporal period of the transmitting waveform. The remainder of this paper is organized as follows: Section 2 describes the design of the OISL-based radar system; Section 3 evaluates the performance of the proposed radar system, including microwave photonic generation, photonics-based de-chirping, targets detection and imaging, as well as flexible tunability; Section 4 discusses and summarizes this work.

## 2. RADAR SYSTEM DESIGN

The schematic diagram of the proposed RF-source-free MWP radar based on an OISL is sketched in Fig. 1. In the transmitter, A continuous-wave (CW) light from ML is split into two branches by an 80/20 optical coupler (OC). The optical signal in the lower branch is used as the optical carrier in the receiver. In the upper branch, the optical signal is injected into the slave laser (SL) after passing through a variable optical attenuator (VOA), a dual-drive Mach–Zehnder modulator (DMZM1), and an optical circulator (CIR). Here, the VOA is used to achieve suitable injection strength. Under proper optical injection parameters, i.e., injection strength and frequency detuning between ML and SL, the injected SL can operate in the desired P1 oscillation state [41]. For a given detuning frequency, the P1 frequency would increase with an increasing injection strength. Therefore, a low-speed control signal $V(t)$ is used to drive the MZM for rapid variation of the optical injection strength. By properly setting the $V(t)$, a broadband LFM signal can be generated after the photodetector (PD1) [31]. Then, in order to improve the phase-noise performance of the generated LFM signal, a delay-matched optoelectronic feedback loop is built, where a half of the LFM signal is separated and fed back to drive the DMZM1, and a variable electrical attenuator (VEA) is inserted to optimize the feedback strength [33, 34]. Here, the round-trip time of the feedback loop, which is mainly induced by a section of single-mode fiber (SMF), needs to be carefully matched with the temporal period of the LFM signal to successfully establish Fourier domain mode locking [42]. It is worth noting that a variable electrical attenuator (VEA) is inserted in the optoelectronic feedback loop to optimize the feedback strength. Afterwards, the performance-enhanced LFM signal is amplified by an electrical amplifier (EA1) before split into two parts by an electrical power divider (Div). One part of the LFM signal is used as a reference for de-chirping the radar echoes, and the other part is launched into air through a transmit antenna (Tx) for target detection. The echoes reflected from the targets are collected by a receive antenna (Rx) and properly amplified by another electrical amplifier (EA2). In the receiver, the reference LFM signal and echoes are separately applied to one of the arms of a dual-drive Mach-Zehnder modulator (DMZM2). The modulated optical signal is then sent to another photodetector (PD2) to perform photonic frequency mixing. After PD2, an electrical low-pass filter (LPF) is employed to remove high-frequency interference and a de-chirped signal is generated. Because of the bandwidth compression property brought by broadband photonic de-chirping, fast and even real-time radar signal processing is possible. Subsequently, the de-chirped signal can be sampled by a low-speed ADC with a high precision. Then, a simple spectral analysis is performed by a digital signal processor (DSP) to acquire the range information of targets. For the simplest situation of single-target detection, considering a dual-chirp LFM signal with up- and down-chirp alternatively is adopted, a de-chirped signal with a frequency of $\Delta f = 2B\Delta\tau/T$ will be obtained, where $B$ is the bandwidth, $T$ is the temporal period and $\Delta\tau$ is the time delay. The distance of the target is

$$L = \frac{\Delta\tau}{2}c = \frac{c}{4B}T\Delta f \quad (1)$$

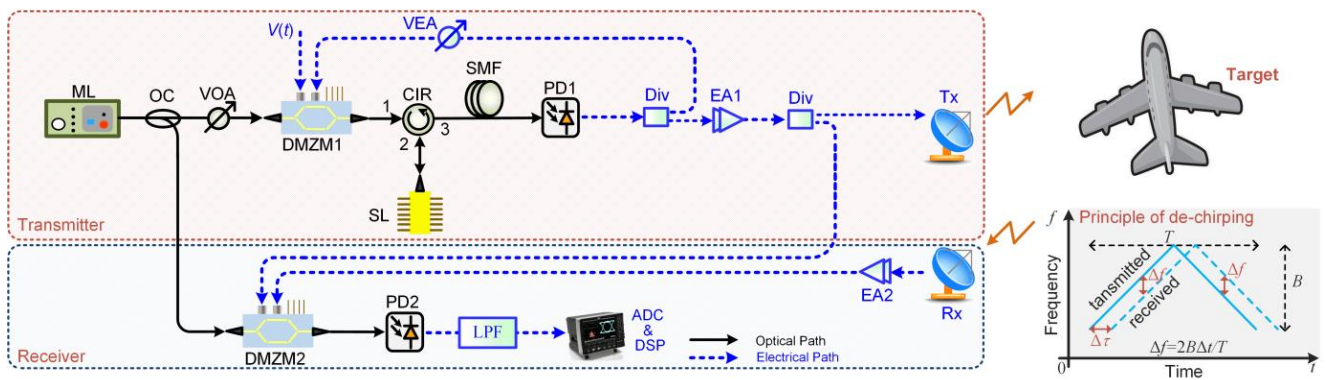

**Fig. 1.** Schematic diagram of the proposed RF-source-free MWP radar. ML, master laser; OC, optical coupler; VOA, variable optical attenuator; DMZM, dual-drive Mach-Zehnder modulator; CIR, optical circulator; SL, slave laser; SMF, single-mode fiber; PD, photodetector; Div, power divider; VEA, variable electrical attenuator; EA, electrical amplifier; LPF, low pass filter; ADC, analog to digital converter; DSP, digital signal processing; Tx, transmitting antenna; Rx, receiving antenna.

The minimum spectral spacing that can be distinguished is $\Delta f_{min} = 1/T$, thus the range resolution is

$$L_{RES} = \frac{c}{4B} \quad (2)$$

## 3. EXPERIMENT AND RESULTS

To verify the feasibility of the proposed RF-source-free MWP radar system based on an OISL, an experiment is performed based on the setup in Fig. 1. A commercial distributed-feedback (DFB) semiconductor laser (Wuhan 69 Inc.) serves as the SL. As shown in Fig. 2(a), its free-running frequency ($f_{SL}$) and optical power are 193.284 THz and 6.5 dBm, respectively. A CW light having a frequency of $f_{ML}$=193.280 THz, namely a detuning of -4 GHz, from the ML (Newkey Photonics Inc.) is injected into the SL with an injection power of -9 dBm, as depicted in the blue curve in Fig. 2 (a). In this case, the SL outputs a typical single-sideband (SSB) P1 optical spectrum. As presented in the gray curve in Fig. 2(a), the sidebands are equally separated by the P1 oscillation frequency $f_0$=16.5 GHz. Two highly dominant components, i.e., a red-shifted cavity mode and a regenerated optical carrier, are observed to be more than 20 dB stronger than the other components. At the output of PD1 (Optilab Inc., 30 GHz), a microwave signal can be generated with a frequency of $f_0$. It has been proved that the microwave frequency $f_0$ can be easily controlled by varying the optical injection strength, which can be achieved by tuning the variable optical attenuator or bias voltage of DMZM1. We measure the relationship between the P1 oscillation frequency and the bias voltage applied to DMZM1 (Fujitsu FTM7937, bandwidth ~25 GHz), as shown in Fig. 2(b). The measured relationship is not ideally linear, which is induced by the nonlinear amplitude transfer function of injected SL and MZM [34, 37].

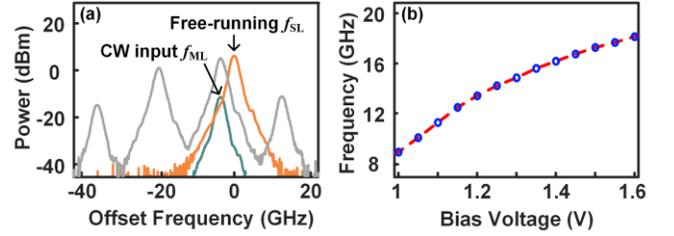

**Fig. 2.** (a) Optical spectra of free-running ML (green curve) and CW input SL (orange curve), P1 dynamics (gray curve) (b) The relationship between the P1 oscillation frequency and the bias voltage applied in DMZM1.

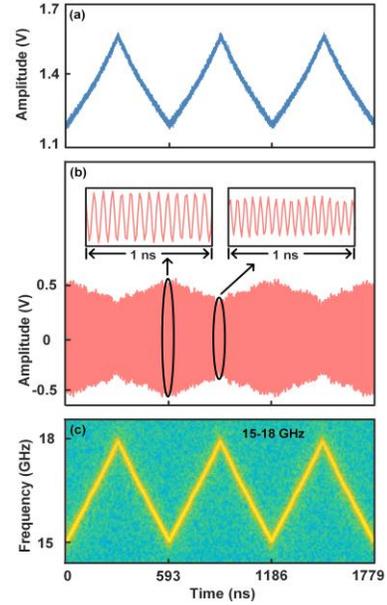

**Fig. 3.** The results for dual-chirp LFM signal generation. (a) Bias voltage applied to DMZM1, (b) temporal waveforms, and (c) time-frequency characteristics.

In order to generate a desired dual-chirp LFM signal for radar detection, the profile of bias voltage $V(t)$ is designed based on the measured relationship in Fig. 2(b). As plotted in Fig. 3(a), a control signal with an amplitude of 0.32 V and a period of 593 ns is applied to DMZM1. Consequently, a dual-chirp LFM signal is acquired in a 20-GHz oscilloscope (LeCroy 820Zi-B), and its temporal waveform is given in Fig. 3(b). The insets show the detailed waveforms at the minimum and maximum instantaneous frequency of a period, respectively. Figure 3(c) depicts the instantaneous frequency characteristics of the acquired waveform based on short-time Fourier transform. As can be seen, a dual-chirp LFM is successfully generated with a bandwidth of 3 GHz (from 15 to 18 GHz) and a period of 593 ns.

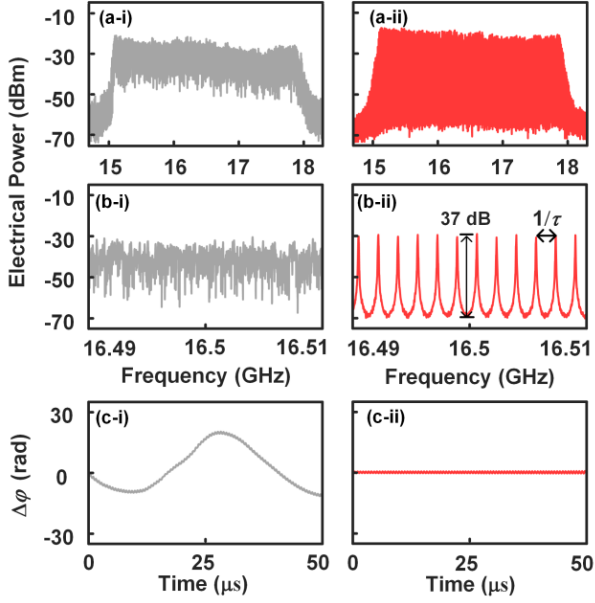

**Fig. 4.** (a) Electrical spectra, (b) detailed electrical spectra, and (c) phase deviation $\Delta\varphi$ of the generated dual-chirp LFM signal. (i) Without feedback, (ii) with feedback.

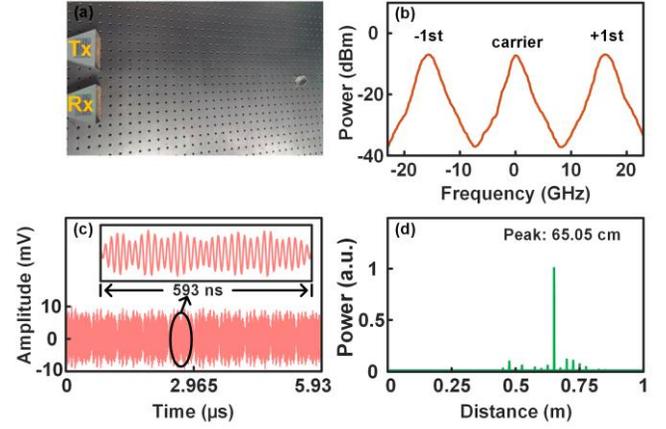

**Fig. 5.** (a) A picture of the antenna pair and target, (b) optical spectrum after DMZM2, (c) temporal waveform and (d) normalized electrical spectrum of the de-chirped signal.

For noise reduction, optoelectronic feedback stabilization is adopted by connecting the generated dual-chirp LFM signal to the RF port of the DMZM1, and the electrical feedback power is around -20 dBm. Figures 4(a) and 4(b) show the electrical spectra of the generated dual-chirp LFM signals (i) with and (ii) without feedback, which is measured by a 40-GHz electric spectrum analyzer (R&S FSV40). When the optoelectronic feedback stabilization is not applied, the generated dual-chirp LFM signal suffers from large phase noise and uncorrelated phase relationship. As shown in Figs. 4(a-i) and 4(b-i), no obvious comb-like features are observed in its electrical spectra. By contrast, when the optoelectronic feedback stabilization is enabled by carefully matching the round-trip time of the feedback loop ($\tau$) to the period of the dual-chirp LFM signal, Fourier domain mode-locking (FDML) state can be established. Due to the fixed phase relationship and reduced phase noise, a comb-like spectrum with a high signal-to-noise ratio (SNR) is obtained in Fig. 4(a-ii). A zoomed spectrum with a frequency span of 20 MHz is plotted in Fig. 4(b-ii), sharp comb components separated by $1/\tau$ are observed, and the comb contrast $R$ reaches 37 dB. The effect of feedback stabilization is further verified by calculating the phase deviation $\Delta\varphi$ of the generated dual-chirp LFM signal. Here, phase deviation $\Delta\varphi$ is calculated by mixing the measured dual-chirp LFM waveform with an ideal one, and the results are plotted in Fig. 4 (c) over a long duration of 50 μs. In Fig. 4 (c-i), the generated dual-chirp LFM signal has a significant phase deviation fluctuation, indicating its poor phase noise performance without feedback. As for Fig. 4 (c-ii), with feedback, the phase deviation fluctuation of the dual-chirp LFM signal is significantly reduced and remains within $\pm\pi/7$ for a duration time of 50 μs. Thus, effective phase noise reduction is achieved through optoelectronic feedback stabilization.

Next, the capability of the photonics-based de-chirping in our radar system is evaluated. The stabilized dual-chirp LFM signal (15-18 GHz, 593ns) is amplified by EA1 (4.5-18.6 GHz, 40 dB) before sent to the transmit antenna (Tx, 12.4 GHz-18 GHz). A trihedral corner reflector (TCR) made of aluminum acts as a target. An individual receive antenna (Rx, 12.4 GHz-18 GHz) is used to receive the echo. The target is placed 65 cm away from the antenna pair, as presented in Fig. 5 (a). After reflected by the target, the echo is firstly amplified by EA2 (6-18.5 GHz, 28 dB) and then connected to the RF port of the DMZM2 (Fujitsu FTM7937) to modulate the optical carrier. The bias voltage of the DMZM2 is set at the minimum transmission point (MITP) because it can suppress the optical carrier to reduce the optical power and eliminates the interference of high-order sidebands. The optical spectrum after DMZM2 is shown in Fig. 5 (b), where the ±1st-order sidebands contain both the local reference signal and the echo signal. The PD2 (Conquer Inc., 10 GHz) is used to detect the frequency difference between the reference signal and the echo signal, and the resultant beating signal passes through an LPF with a bandwidth of 100 MHz to get the de-chirped signal. The de-chirped signal is then captured by an electrical ADC with a sampling rate of 500 MSa/s. Figure 5(c) shows the captured waveform of the de-chirped signal in ten periods, and a detailed waveform of one period is shown in the inset. Based on Equation (1), the normalized electrical spectrum of the de-chirped signal is obtained by performing a fast Fourier transform (FFT) on the captured waveforms. From Fig. 5(d), a dominant spectral peak at 65.05 cm is observed, indicating a small measurement error of only 0.5 mm is achieved.

Then, a dual-target detection experiment is performed, and the system configuration is illustrated in Fig. 6(a), where two metal targets are placed side by side but separated by a distance of $\Delta D$ along the range direction. Figure 6(b) is the power spectrum of the de-chirped signal when the two targets are 35 cm and 60 cm away from the antenna pair, respectively. The two peaks are located at 35.08 cm and 60.05 cm, and the calculated distance ΔD is 24.97 cm. In Fig. 6(c), the distance between the two targets is reduced to 2.5 cm, which is equal to the theoretical range resolution, two spectral peaks with a spacing of 2.5 cm are still observed in Fig. 6(d), indicating that these two targets can be easily distinguished.

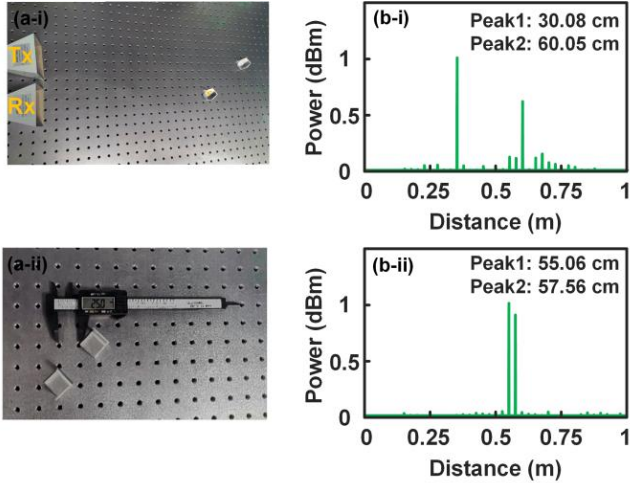

**Fig. 6.** (a) Configuration for dual-target detection (b) normalized electrical spectrum of the de-chirped signal. (i) When two targets are separated by 25 cm, (ii) When two targets are separated by 2.5 cm.

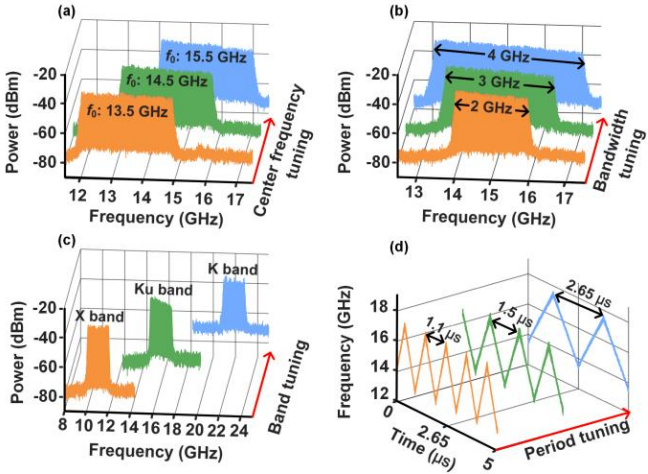

**Fig. 7.** Reconfigurability of the transmitting waveform. (a) The center frequency is tuned from 13.5 GHz to 15.5 GHz, (b) the bandwidth is tuned from 2GHz to 4 GHz, (c) the frequency band is tuned from X band to Ku band, and (d) the temporal period is tuned from 1.1 μs to 2.65 μs.

Multifunction adaptive radar with frequency agility and other reconfigurable operating parameters is highly desired in future radar applications. Since the proposed RF-source-free radar has eliminated the frequency limitation imposed by external RF sources, it can be easily reconfigured to other operating parameters. As shown in Fig. 7, the flexible tunability of the radar is also investigated by adjusting the central frequency, bandwidth, frequency band, and temporal period of the transmitting waveform. For central frequency tuning, the injection power of the ML is carefully tuned while other parameters remain constant. As can be seen in Fig. 7 (a), the central frequency of the electrical spectra is tuned to 13.5 GHz, 14.5 GHz, and 15.5 GHz in a bandwidth of 3 GHz by changing the optical injection power from -12 dBm, to -11 dBm, and -10 dBm. For bandwidth tuning, the amplitude of bias voltage $V(t)$ applied to DMZM1 is tuned while other parameters remain fixed. Figure 7 (b) shows the electrical spectra of the generated dual-chirp LFM signals with different bandwidths, where the bandwidth of 2 GHz, 3 GHz, and 4 GHz corresponds to the amplitude of 0.22 V, 0.32 V, and 0.4 V, respectively. The capability of operating in different frequency bands is demonstrated in Fig. 7(c). By setting the master-slave detuning frequency to be -8 GHz, -4 GHz, and 15 GHz, respectively, the dual-chirp LFM signal is successfully generated in X, Ku, and K bands. In addition, tuning of the temporal period is realized by adjusting the period of $V(t)$ and matching the round-trip time of the feedback loop with different lengths of SMF. As can be seen in Fig. 7(d), the time-frequency curves of the generated dual-chirp LFM signals (13.5-17.5 GHz) with a period of 1.1 μs, 1.5 μs, and 2.65 μs are respectively plotted based on Hilbert transform. It should be noted that the central frequency and bandwidth of the dual-chirp LFM signal by the developed radar transmitter can be further increased if a PD with a larger bandwidth is available.

Next, a proof-of-concept microwave ISAR imaging experiment where the proposed MWP radar operates in Ku band with a bandwidth of 4 GHz (13-17 GHz) and a temporal period of 2.65 μs is conducted. As shown in Fig. 8(a), the ranging resolution is analyzed by performing a dual-target detection experiment when the two targets are placed side by side but separated by a distance of 1.88 cm along the range direction. After a 100 MSa/s ADC and DSP, the result of photonics-based de-chirping is plotted in Fig. 8(b), where two dominant peaks are located at 74.16 cm and 76.02 cm, resulting in a calculated distance of 1.86 cm. The calculated result is very close to the actual value, indicating that the range resolution reaches ~1.88 cm due to the increased bandwidth of 4 GHz.

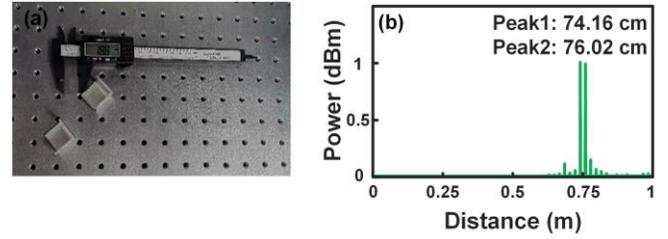

**Fig. 8.** (a) Configuration for dual-target detection, (b) normalized electrical spectrum of the de-chirped signal when two targets are separated by 1.88 cm.

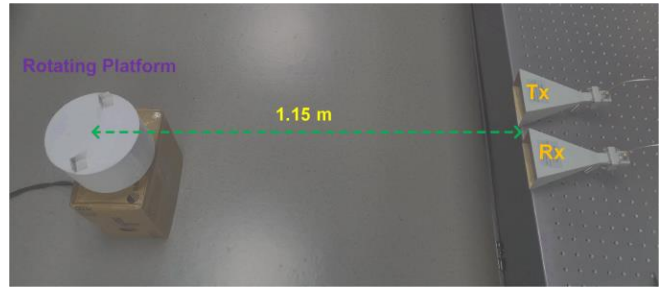

**Fig. 9.** Photograph of the experimental setup including an antenna pair and a rotating platform.

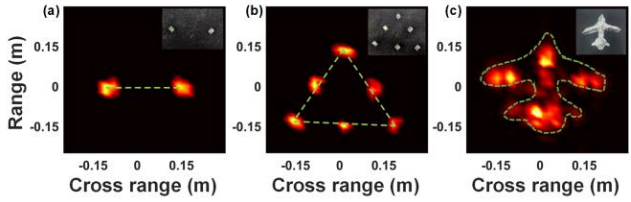

**Fig. 10.** ISAR images of (a) two targets, (b) multiple targets, and (c) an airplane model.

Figure 9 is the photograph of the experimental setup of ISAR imaging including an antenna pair and a rotating platform. The rotating platform is placed at a distance of 115 cm away from the antenna pair, and is set to have a rotation speed of 900° per second. The cross-range resolution of ISAR imaging is given by [9]

$$C_{RES} = \frac{c}{2\theta f_c} \quad (3)$$

where $f_c$ is the center frequency of the dual-chirp LFM signal, $\theta$ is the angle of target rotating during the coherent integration time. Figure 10 is the ISAR image results, in which the pictures of the actual object are illustrated as the insets. The ISAR images are constructed by executing a Range-Doppler (RD) algorithm on the de-chirped signal in an integration time of 31.8 ms. The corresponding ISAR viewing angle $\theta$ and cross-range resolution $C_{RES}$ is 0.5 rad and ~2.00 cm, respectively. According to Fig. 8, the range resolution reaches ~1.88 cm, leading to a 2D imaging resolution as high as ~1.88 cm × ~2.00 cm. Figures 10(a)-10(c) show the constructed ISAR images. First, two targets are placed on the rotating platform with a cross-range distance of 30 cm. Figure 10(a) shows the corresponding imaging results, in which the two targets can be clearly distinguished, and the measured distance agrees well with the real value. Then, ISAR imaging of multiple targets like an equilateral triangle constitute by six TCRs is performed. The results in Fig. 10(b) confirm the high resolution of the proposed MWP radar base on an OISL. Finally, the ISAR imaging of an airplane model (size: 30 cm × 30 cm) is also obtained, as shown in Fig. 10(c), which again clearly proves that a high-resolution imaging radar is accomplished.

## 4. CONCLUSION AND DISCUSSION

In the current study, we proposed and demonstrated a novel microwave photonics radar based on controlled period-one dynamics of an optically injected semiconductor laser. Compared with existing radar systems, the proposed MWP radar is capable of optically generating and processing broadband LFM microwave signals without using any RF sources. By taking advantage of an optically injected semiconductor laser for broadband LFM signal generation and a DMZM-based photonic frequency mixer for the de-chirping process, fast and high-resolution target detection is experimentally demonstrated. In the experiment, a fully photonics-based Ku-band radar with a bandwidth of 4 GHz (13-17 GHz, 2.65 μs) is established for high-resolution detection and ISAR imaging. Results show that a high range resolution reaching 1.88 cm, and a 2D imaging resolution as high as ~1.88 cm × ~2.00 cm are achieved with a sampling rate of 100 MSa/s in the receiver. Besides, the flexible tunability of the radar is also investigated by adjusting the central frequency, bandwidth, frequency band, and temporal period of the transmitting waveform. Thanks to the broadly tunable P1 oscillation frequency of an OISL, the central frequency and bandwidth of the developed radar scheme can be further increased, if PDs, antennas, and electrical amplifiers with a larger bandwidth are available. The proposed radar scheme features low cost, simple structure, and high reconfigurability, which, hopefully, is to be used in future multifunction adaptive and miniaturized radars.

**Funding.** National Natural Science Foundation of China (62001317, 62004135); Natural Science Foundation of Jiangsu Province (BK20200855); Natural Science Research Project of Jiangsu Higher Education Institutions of China (20KJB510011, 20KJA416001); Open Fund of State Key Laboratory of Information Photonics and Optical Communications (Beijing University of Posts and Telecommunications), P. R. China (IPOC2020A012); State Key Laboratory of Advanced Optical Communication Systems Networks, China (2021GZKF003); Project of Key Laboratory of Radar Imaging and Microwave Photonics (Nanjing University of Aeronautics and Astronautics), Ministry of Education (RIMP2020001); Startup Funding of Soochow University (Q415900119).

**Acknowledgments.** The authors would like to thank Qingshui Guo of Zhejiang Lab for helpful discussions and valuable suggestions.

**Disclosures.** The authors declare no conflicts of interest.

†These authors contributed equally to this work.

## References

1. M. I. Skolnik, "An introduction and overview of radar," in *Radar Handbook* (McGraw-Hill, 2008), pp. 1.1-1.24.
2. B. B. Cheng, G. Jing, C. Wang, C. Yang, Y. W. Cai, Q. Chen, X. Huang, G. H. Zeng, J. Jiang, X. J. Deng, and J. Zhang, "Real-time imaging with a 140 GHz inverse synthetic aperture radar," IEEE Trans. THz Sci. Technol. **3**, 606–616 (2013).
3. S. Pan and Y. Zhang, "Microwave Photonic Radars," J. Lightw. Technol. **38**, 5450-5484 (2020).
4. J. D. McKinney, "Photonics illuminates the future of radar," Nature **507**, 310-314 (2014).
5. P. Ghelfi, F. Laghezza, F. Scotti, G. Serafino, A. Capria, S. Pinna, D. Onori, C. Porzi, M. Scaffardi, A. Malacarne, V. Vercesi, E. Lazzeri, F. Berizzi, and A. Bogoni, "A fully photonics-based coherent radar system," Nature **507**, 341-345 (2014).
6. N. Qian, W. Zou, S. Zhang, and J. Chen, "Signal-to-noise ratio improvement of photonic time-stretch coherent radar enabling high-sensitivity ultrabroad W-band operation," Opt. Lett. **43**, 5869-5872 (2018).
7. S. Peng, S. Li, X. Xue, X. Xiao, D. Wu, X. Zheng, and B. Zhou, "High-resolution W-band ISAR imaging system utilizing a logic-operation-based photonic digital-to-analog converter," Opt. Express **26**, 1978–1987 (2018).
8. R. Li, W. Li, M. Ding, Z. Wen, Y. Li, L. Zhou, S. Yu, T. Xing, B. Gao, Y. Luan, Y. Zhu, P. Guo, Y. Tian, and X. Liang, "Demonstration of a microwave photonic synthetic aperture radar based on photonic-assisted signal generation and stretch processing," Opt. Express **25**, 14334–14340 (2017).

9. F. Zhang, Q. Guo, S. Pan, "Photonics-based real-time ultra-high-range-resolution radar with broadband signal generation and processing," Sci. Rep. **7**, 13848 (2017).
10. A. Wang, J. Wo, X. Luo, Y. Wang, W. Cong, P. Du, J. Zhang, B. Zhao, J. Zhang, Y. Zhu, J. Lan, and L. Yu, "Ka-band microwave photonic ultra-wideband imaging radar for capturing quantitative target information," Opt. Express **26**, 20708–20717 (2018).
11. S. Li, Z. Cui, X. Ye, J. Feng, Y. Yang, Z. He, R. Cong, D. Zhu, F. Zhang, and S. Pan, "Chip-based microwave-photonic radar for high resolution imaging," Laser Photonics Rev. **14**, 1900239 (2020).
12. X. Ye, F. Zhang, Y. Yang, and S. Pan, "Photonics-based radar with balanced I/Q de-chirping for interference-suppressed high-resolution detection and imaging," Photonics Res. **7**, 265–272 (2019).
13. F. Zhang, B. Gao, and S. Pan, "Photonics-based MIMO radar with high-resolution and fast detection capability," Opt. Express **26**, 17529–17540 (2018).
14. F. Scotti, S. Maresca, L. Lembo, G. Serafino, A. Bogoni, and P. Ghelfi, "Widely distributed photonics-based dual-band MIMO radar for harbour surveillance," Photonics Technol. Lett. **32**, 1081–1084 (2020).
15. P. Ghelfi, F. Laghezza, F. Scotti, D. Onori, and A. Bogoni, "Photonics for radars operating on multiple coherent bands," J. Lightwave Technol. **34**, 500–507 (2016).
16. J. Cao, R. Li, J. Yang, Z. Mo, J. Dong, X. Zhang, W. Jiang, and W. Li, "Photonic deramp receiver for dual-band LFM-CW radar," J. Lightwave Technol. **37**, 2403–2408 (2019).
17. B. Gao, F. Zhang, E. Zhao, D. Zhang, and S. Pan, "High-resolution phased array radar imaging by photonics-based broadband digital beamforming," Opt. Express **27**, 13194–13203 (2019).
18. A. Wang, D. Zheng, S. Du, J. Zhang, P. Du, Y. Zhu, W. Cong, X. Luo, Y. Wang, and Y. Dai, "microwave photonic radar system with ultra-flexible frequency-domain tunability," Opt. Express **29**, 13887–13898 (2021).
19. X. Zhang, H. Zeng, J. Yang, Z. Yin, Q. Sun, and W. Li, "Novel RF-source-free reconfigurable microwave photonic radar," Opt. Express **28**, 13650-13661 (2020).
20. X. Q. Qi, and J. M. Liu, "Photonic microwave applications of the dynamics of semiconductor lasers," IEEE J. Sel. Topics Quantum Electron. **10**, 1198–1211 (2011).
21. S. C. Chan, S. K. Hwang, and J. M. Liu, "Radio-over-fiber AM-to-FM upconversion using an optically injected semiconductor laser," Opt. Lett. **31**, 2254–2256 (2006).
22. S. C. Chan, and J. M. Liu, "Tunable narrow-linewidth photonic microwave generation using semiconductor laser dynamics," IEEE J. Sel. Top. Quantum Electron. **10**, 1025-1032 (2004).
23. Y. H. Hung, and S. K. Hwang, "Photonic microwave stabilization for period-one nonlinear dynamics of semiconductor lasers using optical modulation sideband injection locking," Opt. Express **23**, 6520-6532 (2015).
24. J. S. Suelzer, T. B. Simpson, P. Devgan, and N. G. Usechak, "Tunable, low-phase-noise microwave signals from an optically injected semiconductor laser with opto-electronic feedback," Opt. Lett. **42**, 3181-3184 (2017).
25. P. Zhou, F. Z. Zhang, D. C. Zhang, and S. L. Pan, "Performance enhancement of an optically-injected-semiconductor-laser-based optoelectronic oscillator by subharmonic microwave modulation," Opt. Lett. **43**, 5439-5442 (2018).
26. Y. Li, L. Fan, G. Xia, and Z. Wu, "Tunable and broadband microwave frequency comb generation using optically injected semiconductor laser nonlinear dynamics," IEEE Photonics J., **9**, 1-7 (2017).
27. H. Zhu, R. Wang, T. Pu, P. Xiang, J. Zheng, and T. Fang, "A novel approach for generating flat optical frequency comb based on externally injected gain-switching distributed feedback semiconductor laser," Laser Phys. Lett. **14**, 02601 (2017).
28. Y. Doumbia, T. Malica, D. Wolfersberger, K. Panajotov, and M. Sciamanna, "Nonlinear dynamics of a laser diode with an injection of an optical frequency comb," Opt. Express **28**, 30379-30390 (2020).
29. P. Zhou, F. Zhang, B. Gao, and S. Pan, "Optical pulse generation by an optoelectronic oscillator with optically injected semiconductor laser," IEEE Photonics Technol. Lett. **28**, 1827–1830 (2016).
30. P. Zhou, F. Zhang, Q. Guo, and S. Pan, "A modulator-free photonic triangular pulse generator based on semiconductor lasers," IEEE Photonics Technol. Lett. **30**, 1317–1320 (2018).
31. P. Zhou, F. Zhang, Q. Guo, and S. Pan, "Linearly chirped microwave waveform generation with large time-bandwidth product by optically injected semiconductor laser," Opt. Express **24**, 18460–18467 (2016).
32. P. Zhou, F. Zhang, Q. Guo, and S. Pan, "Reconfigurable radar waveform generation based on an optically injected semiconductor laser," IEEE J. Sel. Top. Quantum Electron. **23**, 1801109 (2017).
33. J. P. Zhuang, X. Z. Li, S. S. Li, and S. C. Chan, "Frequency-modulated microwave generation with feedback stabilization using an optically injected semiconductor laser," Opt. Lett. **41**, 5764–5767 (2016).
34. P. Zhou, F. Z. Zhang, and S. L. Pan, "Generation of linearly chirped microwave waveform with an increased time-bandwidth product based on a tunable optoelectronic oscillator," J. Lightwave Technol. **36**, 3726–3732 (2018).
35. X. D. Lin, G. Q. Xia, Z. Shang, T. Deng, X. Tang, L. Fan, Z. Y. Gao, and Z. M. Wu, "Frequency-modulated continuous-wave generation based on an optically injected semiconductor laser with optical feedback stabilization," Opt. Express **27**, 1217–1225 (2019).
36. P. Zhou, F. Zhang, X. Ye, Q. Guo, and S. Pan, "Flexible frequency-hopping microwave generation by dynamic control of optically injected semiconductor laser," IEEE Photonics J. **8**, 5501909 (2016).
37. P. Zhou, R. Zhang, K. Li, Z. Jiang, P. Mu, H. Bao, and N. Li, "Generation of NLFM microwave waveforms based on controlled period-one dynamics of semiconductor lasers," Opt. Express **28**, 32647-32656 (2020).
38. B. Zhang, D. Zhu, P. Zhou, C. Xie, and S. Pan, "Tunable triangular frequency modulated microwave waveform generation with improved linearity using an optically injected semiconductor laser," Appl. Opt. **58**, 5479-5485 (2019).
39. P. Zhou, H. Chen, N. Li, R. Zhang, and S. Pan, "Photonic generation of tunable dual-chirp microwave waveforms using a dual-beam optically injected semiconductor laser," Opt. Lett. **45**, 1342-1345 (2020).
40. P. Zhou, R. H. Zhang, Z. D. Jiang, N. Q. Li and S. L. Pan, "Demonstration of a RF-source-free microwave photonic radar based on an optically injected semiconductor laser," in *Optical Fiber Communications Conference and Exposition (OFC)*, Washington, DC (OSA, 2021), paper Th1A.11.
41. S. K. Hwang, J. M. Liu, and J. K. White, "Characteristics of period-one oscillations in semiconductor lasers subject to optical injection," IEEE J. Sel. Top. Quantum Electron. **10**, 974–981 (2004).
42. T. Hao, Q. Cen, Y. Dai, J. Tang, W. Li, J. Yao, N. Zhu, and M. Li, "Breaking the limitation of mode building time in an optoelectronic oscillator," Nat. Commun. **9**, 1839 (2018).